\documentclass{article}
\usepackage{ijcai17}
\usepackage{graphicx}
\usepackage{amsfonts}
\usepackage{multirow}
\usepackage{amsmath}

\usepackage{times}

\title{Modeling Text with Graph Convolutional Network for \\Cross-Modal Information Retrieval}

\author{Jing Yu$^1$, Yuhang Lu$^{1,2}$, Zengchang Qin$^3$, Yanbing Liu$^1$, Jianlong Tan$^1$, Li Guo$^1$, Weifeng Zhang$^4$\\
$^1$Institute of Information Engineering, Chinese Academy of Sciences, China  \\
$^2$School of Cyber Security, University of Chinese Academy of Sciences, China  \\
$^3$Intelligent Computing \& Machine Learning Lab, School of ASEE, Beihang University, China  \\
$^4$Hangzhou Dianzi University, China  \\
\{yujing02, luyuhang, liuyanbing, tanjianlong, guoli\}@iie.ac.cn, zcqin@buaa.edu.cn, zwf.zhang@gmail.com}

\begin{document}

\maketitle

\begin{abstract}
Cross-modal information retrieval aims to find heterogeneous data of various modalities from a given query of one modality.
The main challenge is to map different modalities into a common semantic space, in which distance between concepts in different modalities can be well modeled.
For cross-modal information retrieval between images and texts,
existing work mostly uses off-the-shelf Convolutional Neural Network (CNN) for image feature extraction. For texts, word-level features such as bag-of-words
or word2vec are employed to build deep learning models to represent texts.
Besides word-level semantics, the semantic relations between words are also informative
but less explored.
In this paper, we model texts by graphs using similarity measure based on word2vec.
A dual-path neural network model is proposed for couple feature learning in cross-modal information retrieval.
One path utilizes Graph Convolutional Network (GCN) for text modeling based on graph representations. The other path uses
a neural network with layers of nonlinearities for image modeling based on off-the-shelf features.
The model is trained by a pairwise similarity loss function to maximize the similarity of relevant text-image pairs and minimize the similarity of irrelevant pairs.
Experimental results show that the proposed model outperforms the state-of-the-art methods significantly, with 17\% improvement on accuracy for the best case.

\end{abstract}

\section{Introduction}
\label{sec:introduction}

\begin{figure}[!h]
\begin{minipage}{1\linewidth}
  \centerline{\includegraphics[width=8.5cm,height = 2.9cm]{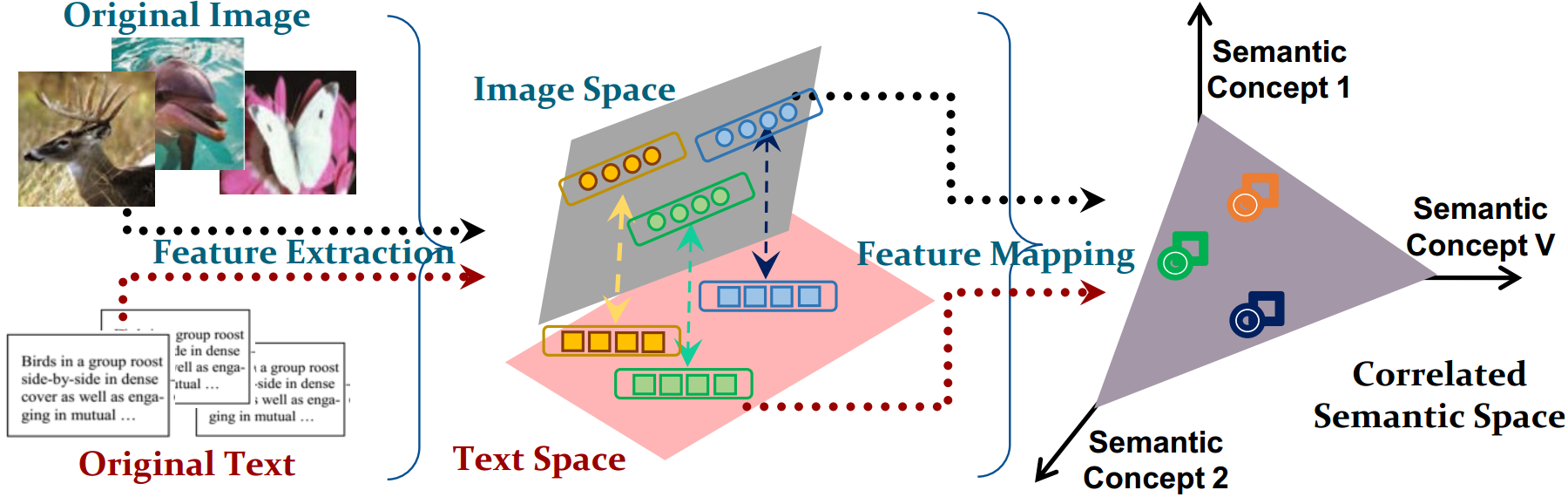}}
  \centerline{\scriptsize{(a) Overview of classical cross-modal retrieval models.}}
\end{minipage}

\begin{minipage}{1\linewidth}
  \centerline{\includegraphics[width=8.5cm,height = 3.2cm]{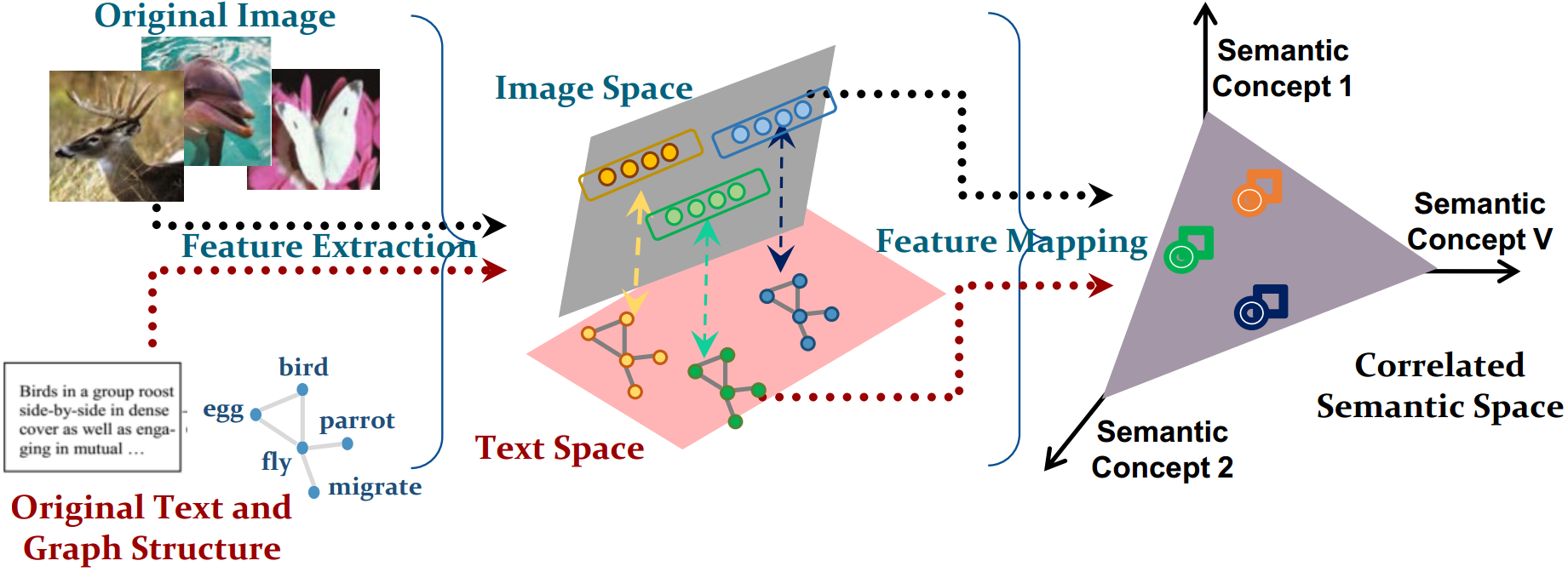}}
  \centerline{\scriptsize{(b) Overveiw of our proposed cross-modal retrieval model.}}
\end{minipage}
\caption{Comparison of classical cross-modal retrieval models to our model. (a) Classical models adopt feature vectors to represent grid-structured multimodal data; (b) Our model can handle both irregular graph-structured data and regular grid-structured data simultaneously.}
\label{fig: comparison}
\end{figure}

For past a few decades, online multimedia information in different modalities, such as image, text,
video and audio, has been increasing and accumulated explosively. Information
related to the same content or topic may exist in various modalities and has heterogeneous
properties, that makes it difficult for traditional uni-modal information retrieval
systems to acquire comprehensive information. There is a growing demand for effective and efficient search
in the data across different modalities. Cross-modal information
retrieval \cite{Rasi:cross,Yu:TCM,Wang:joint} enables users to take a query of one modality to retrieve data in relevant
content in other modalities. 
However, there is no natural correspondence between different modalities.
Previous research has made continuous effort on designing appropriate distance measure of similarity and gained
great progress. A common solution is to learn a common latent semantic space to
compare all modalities of data directly, typically using probabilistic models \cite{Blei:model},
metric learning \cite{Zheng:hash}, subspace learning \cite{Hardoon:CCA,Rasi:cross,sharma:multiview}, and joint modeling methods \cite{Wang:joint}.
A brief survey is available in \cite{Wang:joint}.

Feature representation is the footstone for cross-modal information retrieval. In the case of text-image retrieval,
off-the-shelf features learnt by deep models are widely used to represent images.
Most methods \cite{liang:adaptive,Wang:deep} use Convolutional Neural Network (CNN) \cite{Lecun:CNN} to learn the visual features obtained from the pre-trained model for object recognition on ImageNet. CNN can effectively
extract hierarchies of visual feature vectors and the fixed CNN feature can be directly used for text-image
semantic space mapping. In this paper, we also employ the same routine to use
pre-trained CNN for visual feature learning.
For text representation, the popular $vector$-$space$ model is usually adopted to convert a text to a textual vector for learning high-level semantic representations.
In this kind of models, $bag$-$of$-$words$ (BOW) is commonly used in cross-modal information retrieval \cite{Liu:matrix,liang:group}.
Intuitively, the text document is represented by a word-frequency vector regardless of the word order. Although some weighting schemes based on word
frequency have been proposed to enhance the feature discrimination \cite{Wang:joint}, one common problem is that the relations
among words are not considered.
Recently, $word2vec$ \cite{Mikolov:word2vec} becomes one of the best models for semantical modeling of word semantics. It's pre-trained
on GoogleNews to learn the vector representation from the context information. \cite{Wang:deep} extracts word vectors via $word2vec$
model and adopts Fisher vector encoding to obtain the sentence representation. \cite{liang:adaptive} represents a text by
calculating a mean vector of all the word $word2vec$ vectors in a text. Although this kind of word vector is enriched by learning from
neighboring words, it still ignores the global structural information inherent in the texts and only treat the word as ``flat'' features.
In light of the common weakness in $vector$-$space$ models, recent research has found
that the relations among words could provide rich semantics of the texts and can
effectively promote the text classification performance \cite{Wang:HIN}.

In this paper, we represent a text as a structured and featured graph and learn text features by a
graph-based deep model, i.e. Graph Convolutional Network (GCN) \cite{kipf:semi,Deff:fast}.
Such a graph can well capture the semantic relations among words.
GCN allows convolutions to be dealt as multiplication
in the graph spectral domain, rendering the extension of CNN to irregular graphs.
(Figure \ref{fig: comparison} shows the comparison of our model to classical cross-modal retrieval models.)
The GCN model has a great ability to learn local and stationary features on graphs, which was
successfully used in text categorization \cite{kipf:semi} and brain network matching \cite{ktena:metric}.
Based on this graph representation for texts, we propose a dual-path neural network, called \textbf{G}raph-\textbf{I}n-\textbf{N}etwork (\textbf{GIN}), for cross-modal information retrieval.
The text modeling path contains a GCN on the top of graph representations. The image modeling path contains a neural network
with layers of nonlinearities on the top of off-the-shelf image representations.
To train the model, we employ a pairwise similarity loss function \cite{Kumar:loss},
that maximizes the similarity between samples in the same semantic concept
and minimizes the similarity between samples in different semantic concepts.

The main contributions can be summarized as follows:
\begin{itemize}
\item We propose a cross-modal retrieval model to model text data by graph convolutional network, which realizes the cross-modal retrieval between irregular
graph-structured data and regular grid-structured data;
\item The model can jointly learn the textual and visual representations as well as text-image similarity metric, providing an end-to-end training mode;
\item Experimental results on five benchmark datasets show the superior performance of our model over the state-of-the-art methods, verifying the benefits of using graphs to model the irregular textual data.
\end{itemize}

\begin{figure*}[!ht]
\centering
\includegraphics[width = 18cm,height = 7.5cm]{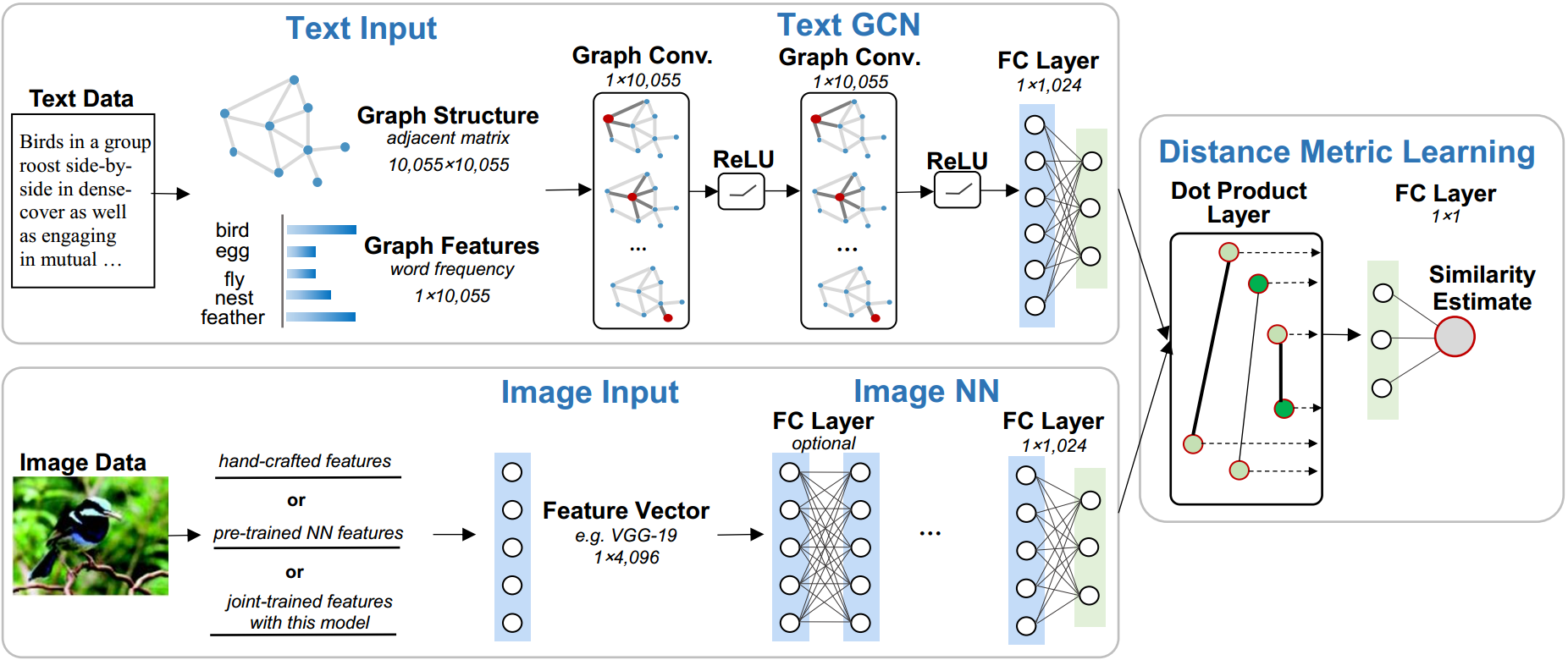}
\caption{The structure of the proposed model is a dual-path neural network: i.e., text Graph Convolutional Network
(text GCN) (top) and image Neural Network (image NN) (bottom). The text GCN for learning text representation contains two
layers of graph convolution on the top of constructed featured graph. The image NN for learning image representation contains
layers of non-linearities initialized by off-the-shelf features. They have the same dimension in the last fully connected
layers. The objective is a global pairwise similarity loss function.
}
\label{fig:framework}
\end{figure*}

\section{Methodology}
\label{sec:method}

In this paper, we propose a dual-path neural network to simultaneously learn multi-modal representations and similarity metric in an end-to-end mode.
\label{subsec:text}
In the text modeling path (top in Figure \ref{fig:framework}, that the convolution part is referred to the blog of GCNs  \footnote{http://tkipf.github.io/graph-convolutional-networks/}), each text is represented by a featured graph and the text GCN is used to learn the feature representation.
It has two key steps: \textit{graph construction} and \textit{GCN modeling}.
\subsection{Text Modeling}
\textbf{Graph Construction:} Classical methods semantically model the fundamental features of a text only by word vectors regardless of the structural information.
In this work, we represent a text by a featured graph to combine the strengths of structural information with semantic information together.
Given a set of text documents, we extract the most common words, denoted as $W=[w_1, w_2, ..., w_N]$, from all the unique words in this corpus and represent each word by a pre-trained $word2vec$ embedding. For the graph structure, we construct a $k$-nearest neighbor graph, denoted as $G=(V,E)$. Each vertex $v_i\in V$ is corresponding to a unique word and each edge
$e_{ij}\in E$ is defined by the $word2vec$  similarity between two words:
\begin{equation}
e_{ij}=\begin{cases}
1 & \text{ if } w_i\in N_k(w_j)\  \text{or} \  w_j\in N_k(w_i)\\
0 & \text{ otherwise }
\end{cases}
\end{equation}
where $N_k(\cdot )$ denotes the set of $k$-nearest neighbors by computing the cosine similarity between word \textit{word2vec} embeddings.
$k$ is the parameter of neighbor numbers (set to 8 in our following experiments). The graph structure is stored by an adjacent matrix $A\in \mathbb{R}^{N\times N}$.
For the graph features, each text document is represented by a
$bag$-$of$-$words$ vector and the frequency value of word $w_i$ serves as the 1-dimensional feature on vertex $v_i$. In this
way, we combine structural information of word similarity relations and semantic information of word vector representation in
a featured graph. Note that the graph structure is identical for a corpus and we use different graph features to represent each text
in a corpus. \\
%
%
\textbf{GCN Modeling:} In modeling text corpora, deep network models have become
increasingly popular and achieved breakthroughs in many machine learning areas.
However, classical deep network models are defined for grid-structured data and can
not be easily extended to graphs. It's challenging to define the local neighborhood structures
and the vertex orders for graph operations. Recently, Graph Convolutional Network (GCN)
is proposed to generalize Convolutional Neural Network (CNN) to irregular-structured graphs.
The basic idea is that, based on spectral graph theory, the graph convolutions
can be dealt as multiplications in the graph spectral domain. The feature maps can be
obtained by inverse transform from the graph spectral domain to original graph domain. In this paper, the
text features are learnt by GCN given the graph representation of a text document.

Given a text, we define its input graph feature vector by $F_{in}$ and we denote the output feature vector after graph
convolution by $F_{out}$. Firstly $F_{in}$ is transformed to the spectral
domain via graph Fourier transform. This transform is based on the normalized
graph Laplacian, defined as $L=I_{N}-D^{-1/2}AD^{-1/2}$, where $I_{N}$ and $D$ are respectively the
identity matrix and diagonal degree matrix of the graph structure $G$. Then $L$ can be eigendecomposed
as $L=U\Lambda U^{T}$, where $U$ is a set of eigenvectors and $\Lambda$ is a set of real, non-negative eigenvalues.
The Fourier transform of $F_{in}$ is a function of $U$ defined as:

\begin{equation}
\widehat{F_{in}}=U^{T}F_{in}
\end{equation}
While the inverse transform is defined as:
\begin{equation}
F_{in}=U\widehat{F_{in}}
\end{equation}
The convolution of $F_{in}$ with a spectral filter $g_{\theta}$ is given by:
\begin{equation}
\label{eq:laplacian}
F_{out}=g_{\theta}\ast F_{in}=Ug_{\theta}U^{T}F_{in}
\end{equation}
where parameter $\theta$ is a vector to learn. In order to keep the filter $K$-localized in space and computationally efficient,
\cite{Deff:fast} proposes a approximated polynomial filter defined as:

\begin{equation}
\label{eq:laplacian}
g_\theta=\sum_{k=0}^{K-1}\theta_kT_k(\widetilde{L})
\end{equation}
where $T_k(x)=2xT_{k-1}(x)-T_{k-2}(x)$ with $T_0(x)=1$ and $T_1(x)=x$, $\widetilde{L}=\frac{2}{\lambda_{max}}L-I_N$
and $\lambda_{max}$ denotes the largest eigenvalue of $L$. The filtering operation can then be written as $F_{out}=g_\theta F_{in}$.
In our model, we use the same filter as in \cite{Deff:fast}. For the graph representation of a text document, the $i^{th}$ input graph feature
$f_{in,i}\in F_{in}$ is the word frequency of vertex $v_i$. Then the $i^{th}$ output feature $f_{out,i}\in F_{out}$ is given by:

\begin{equation}
\label{eq:laplacian}
f_{out,i}=\sum_{k=0}^{K-1}\theta_kT_k(\widetilde{L})f_{in,i}
\end{equation}
where we set $K$$=$$3$ in the experiments to keep each convolution at most 3-steps away from a center vertex.

Our text GCN contains two layers of graph convolutions, each followed by Rectified Linear Unit (ReLU) activation to increase non-linearity.
A fully connected layer is successive with the last convolution layer to map the text features to the common latent semantic space.
Given a text document $T$, the text representation
$f_{t}$ learnt by the text GCN model $H_{t}(\cdot )$ is denoted by:

\begin{equation}
f_{t}=H_{t}(T)
\end{equation}

\subsection{Image Modeling}
\label{subsec:image}
For modeling images, we adopt a neural network (NN) containing a set of fully connected layers (bottom in Figure \ref{fig:framework}).
We have three options of initializing inputs by
hand-crafted feature descriptors, pre-trained neural networks, or jointly trained end-to-end neural networks.
In this paper, the first two kinds of features are used for fair comparison with other models.
The input visual features are followed by a set
of fully connected layers for fine-tuning the visual features. Similar to text modeling, the last fully connected
layer of image NN maps the visual features to the common latent semantic space with the same dimension as text.
In experimental studies, we tune the number of layers and find that only keeping the last semantic mapping layer
without feature fine-tuning layers can obtain satisfactory results. Given an image $I$, the image representation $f_{img}$ learnt by the model
from image NN $H_{img}(\cdot)$ is represented by:

\begin{equation}
f_{img}=H_{img}(I)
\end{equation}

\subsection{Objective Function}
\label{subsec:Loss}
Distance metric learning is applied to estimate the relevance of features learned from the dual-path model. The outputs
of the two paths, i.e. $f_{t}$ and $f_{img}$, are in the same dimension and combined by an inner product layer. The successive layer is a
fully connected layer with one output $score(T,I)$, denoting the similarity score function between a text-image pair.
The training objective is a pairwise similarity loss function proposed in \cite{Kumar:loss}, which outperforms existing works
in the problem of learning local image features. In our research, we maximize
the mean similarity score $u^{+}$ between text-image pairs of the same semantic concept and minimize the mean
similarity score $u^{-}$ between pairs of different semantic concepts. Meanwhile, we also minimises the
variance of pairwise similarity score for both matching $\sigma ^{2+}$ and non-matching $\sigma ^{2-}$ pairs.
The loss function is formally by:
\begin{equation}
\label{eq:loss}
Loss=(\sigma ^{2+}+\sigma ^{2-})+\lambda \max(0,m-(u^{+}-u^{-}))
\end{equation}
where $\lambda$ is used to balance the weight of the mean and variance, and $m$ is the margin between the mean distributions of
matching similarity and non-matching similarity.
$u^{+}$$=$$\sum_{i=1}^{Q_1}\frac{score(T_i,I_i)}{Q_1}$ and $\sigma ^{2+}$$=$$\sum_{i=1}^{Q_1}\frac{(score(T_i,I_i)-u^{+})^{2}}{Q_1}$ when text $T_i$ and image $I_i$ are in the same class. While
$u^{-}$$=$$\sum_{j=1}^{Q_2}\frac{score(T_j,I_j)}{Q_2}$ and $\sigma ^{2-}$$=$$\sum_{j=1}^{Q_2}\frac{(score(T_j,I_j)-u^{-})^{2}}{Q_2}$ when $T_j$ and $I_j$ are in different classes.
 We train the model by mini-batch gradient descent with mini-batch size 200.
In other words, we sequentially select $Q_1+Q_2=200$ text-image pairs from the training set for each mini-batch in the experiments.

\section{Experimental Studies}
To evaluate the performance of our proposed model, we conduct extensive experiments to investigate cross-modal retrieval tasks,
i.e. text-query-images and image-query-texts.

\subsection{Datasets}
\label{subsec: dataset}
Experiments are conducted on four English benchmark datasets, i.e. English Wikipedia, NUS-WIDE, Pascal VOC, and TVGraz. To verify the extensibility
of our model, we also conduct experiments on the Chinese Wikipedia dataset.
Each dataset contains a set of text-image pairs. Images are represented by off-the-shelf feature vectors while texts are represented
by featured graphs. \\
\textbf{English Wikipedia} 
dataset (Eng-Wiki for short)\cite{Rasi:cross} contains 2,866 image-text pairs divided into 10 classes, where 2,173 pairs are for training and 693 pairs are for testing. Each image is represented by a 4,096-dimensional vector extracted from the last fully connected layer of VGG-19 model \cite{Simonyan:VGG19}. Each text is represented by a graph with 10,055 vertices. \\
\textbf{NUS-WIDE} 
dataset consists of 269,648 image-tag pairs, which are pruned from the NUS dataset by keeping the pairs belonging to one or more of the 10 largest classes. We select samples in the  10 largest classes as adopted in \cite{liang:adaptive}. For images, we use 500-dimensional bag-of-features. For tags, we construct a graph with 5,018 vertices.
\\
\textbf{Pascal VOC} 
dataset consists of 9,963 image-tag pairs belonging to 20 classes. The images containing only one object are selected in our
experiments as \cite{sharma:multiview,wang:learn,liang:group}, obtaining 2,808 training and 2,841 testing samples. For the features, 512-dimensional Gist features are adopted for the images and a graph with 598 vertices is used for the tags. \\
\textbf{TVGraz }
dataset contains 2,594 image-text pairs \cite{Costa:role}. We choose the texts that have more than 10 words in our experiments and results in 2,360 image-text pairs, where 1,885 pairs for training and 475 pairs for testing. Each image is represented by a 4,096-dimensional VGG-19 feature  and each text is represented by a graph with 8,172 vertices.
\\
\textbf{Chinese Wikipedia} 
dataset (Ch-Wiki for short) \cite{qin:cross} is collected from Chinese Wikipedia articles. It contains 3,103 image-text pairs divided into 9 classes, where 2,482 pairs are for training and 621 pairs are for testing. Each image is represented by a 4,096-dimensional output of VGG-19. Each text is represented by a graph with 9,613 vertices.


\subsection{Evaluation and Implementation}

We compare our proposed GIN with a number of state-of-the-art models, including CCA \& SCM \cite{Rasi:cross}, TCM \& w-TCM \& c-TCM \cite{qin:cross},
GMLDA \& GMMFA \cite{sharma:multiview}, LCFS \cite{wang:learn}, MvDA \cite{Kan:multi}, LGCFL \cite{Kang:learn}, ml-CCA \cite{Ranjan:multi},
AUSL \cite{liang:adaptive}, JFSSL \cite{Wang:joint}, PLS \cite{sharam:bypassing}, BLM \cite{sharma:multiview}, CDFE \cite{Lin:inter}, CCA-3V \cite{Gong:multi},
CM \& SM \cite{Costa:role}, and TTI \cite{Qi:semantic}. For the same settings with \cite{liang:adaptive}, principal component analysis is performed on the original features
for CCA, SCM, GMLDA and MvDA.

CCA, PLS and BLM are three popular un-supervised models that adopt pairwise information to maximize the
correlation between projected vectors. AUSL and CCA-3V are semi-supervised
models that leverage both labelled and unlabelled data to learn the common space. GMLDA, GMMFA, ml-CCA, TCM,
LCFS, LGCFL, JFSSL, CDFE, and MvDA are supervised models that use the semantic class information to
directly make data from one modality to correlate with data from another modality.

The mean average precision (MAP) is used to evaluate the performance of all the algorithms on the five datasets.
Higher MAP indicates better retrieval performance. Meanwhile, the precision-recall (PR) curve \cite{Rasi:cross} is
also utilized  for evaluation.
In our implementation, we set $k=8$ in $k$-nearest neighbors for text graph construction.
In this work, positive samples denote the image-text pairs
belonging to the same class while negative samples correspond to image-text pairs from different classes. So the ground truth labels are binary
denoting whether the input pairs are from the same class or not. For all the datasets, we randomly select matched and non-matched text-image pairs and form about
40,000 positive samples and 40,000 negative samples for training.
We train the model for 50 epochs with mini-batch size 200. We adopt the dropout ratio of 0.2 at the input of the last FC layer,
learning rate 0.001 with an Adam optimisation, and regularisation 0.005. $m$ and $\lambda$ in the loss function are set to 0.6 and 0.35, respectively.
In the last semantic mapping layers of both text path and image path, the reduced dimensions are set to 1,024, 500, 256, 1,024, 1,024 for Eng-Wiki, NUS-WIDE, Pascal, TVGraz, and Ch-Wiki, respectively. The code will be released on github upon the acceptance of the paper.

\subsection{Experimental Results}
\label{subsec: results}
The MAP scores of all the methods on the five benchmark datasets are shown in Table \ref{tab:MAP}. All the
other models are well cited work in this field. Since not all the papers have tested these five datasets, for fair comparison, we compare our model to methods
on their reported datasets with the same preprocessing conditions. From Table \ref{tab:MAP}, we can have the
following observations:

\begin{table}[!h]
\setlength{\abovecaptionskip}{0pt}
\setlength{\belowcaptionskip}{5pt}
\centering
\caption{MAP score comparison of text-image retrieval on five given benchmark datasets.}
\label{tab:MAP}
\setlength{\tabcolsep}{1mm}
\begin{tabular}{|c|c|c|c|c|}
\hline
\multicolumn{1}{|l|}{\textbf{Method}} & \multicolumn{1}{l|}{\textbf{Text query}} & \multicolumn{1}{l|}{\textbf{Image query}} & \multicolumn{1}{l|}{\textbf{Average}} & \textbf{Dataset}           \\ \hline \hline
CCA                                   & 0.1872                                   & 0.2160                                    & 0.2016                                & \multirow{12}{*}{Eng-Wiki} \\ \cline{1-4}
SCM                                   & 0.2336                                   & 0.2759                                    & 0.2548                                &                            \\ \cline{1-4}
TCM                                   & 0.2930                                   & 0.2320                                    & 0.2660                                &                            \\ \cline{1-4}
LCFS                                  & 0.2043                                   & 0.2711                                    & 0.2377                                &                            \\ \cline{1-4}
MvDA                                  & 0.2319                                   & 0.2971                                    & 0.2645                                &                            \\ \cline{1-4}
LGCFL                                 & 0.3160                                   & 0.3775                                    & 0.3467                                &                            \\ \cline{1-4}
ml-CCA                                & 0.2873                                   & 0.3527                                    & 0.3120                                &                            \\ \cline{1-4}
GMLDA                                 & 0.2885                                   & 0.3159                                    & 0.3022                                &                            \\ \cline{1-4}
GMMFA                                 & 0.2964                                   & 0.3155                                    & 0.3060                                &                            \\ \cline{1-4}
AUSL                                  & 0.3321                                   & 0.3965                                    & 0.3643                                &                            \\ \cline{1-4}
JFSSL                                 & 0.4102                                   & \textbf{0.4670}                           & 0.4386                                &                            \\ \cline{1-4}
GIN                                 & \textbf{0.7672}                          & 0.4526                                    & \textbf{0.6099}                       &                            \\ \hline \hline
CCA                                   & 0.2667                                   & 0.2869                                    & 0.2768                                & \multirow{7}{*}{NUS-WIDE}  \\ \cline{1-4}
LCFS                                  & 0.3363                                   & 0.4742                                    & 0.4053                                &                            \\ \cline{1-4}
LGFCL                                 & 0.3907                                   & 0.4972                                    & 0.4440                                &                            \\ \cline{1-4}
ml-CCA                                & 0.3908                                   & 0.4689                                    & 0.4299                                &                            \\ \cline{1-4}
AUSL                                  & 0.4128                                   & \textbf{0.5690 }                          & 0.4909                                &                            \\ \cline{1-4}
JFSSL                                 & 0.3747                                   & 0.4035                                    & 0.3891                                &                            \\ \cline{1-4}
GIN                                 & \textbf{0.5418}                          & 0.5236                                    & \textbf{0.5327}                       &                            \\ \hline \hline
PLS                                   & 0.1997                                   & 0.2757                                    & 0.2377                                & \multirow{10}{*}{Pascal}   \\ \cline{1-4}
BLM                                   & 0.2408                                   & 0.2667                                    & 0.2538                                &                            \\ \cline{1-4}
CCA                                   & 0.2215                                   & 0.2655                                    & 0.2435                                &                            \\ \cline{1-4}
CDFE                                  & 0.2211                                   & 0.2928                                    & 0.2569                                &                            \\ \cline{1-4}
GMLDA                                 & 0.2448                                   & 0.3094                                    & 0.2771                                &                            \\ \cline{1-4}
GMMFA                                 & 0.2308                                   & 0.3090                                    & 0.2699                                &                            \\ \cline{1-4}
CCA3V                                 & 0.2562                                   & 0.3146                                    & 0.2854                                &                            \\ \cline{1-4}
LCFS                                  & 0.2674                                   & 0.3438                                    & 0.3056                                &                            \\ \cline{1-4}
JFSSL                                 & 0.2801                                   & \textbf{0.3607}                           & 0.3204                                &                            \\ \cline{1-4}
GIN                                 & \textbf{0.4515}                          & 0.3170                                    & \textbf{0.3842}                       &                            \\ \hline \hline
TTI                                  & 0.1530                                   & 0.2160                                    & 0.1845                                &                            \\ \cline{1-4}
CM                                   & 0.4500                                   & 0.4600                                    & 0.4550                                &                            \\ \cline{1-4}
SM                                   & 0.5850                                   & 0.6190                                    & 0.6020                                &
\multirow{3}{*}{TVGraz}    \\ \cline{1-4}
SCM                                   & 0.6960                                   & 0.6930                                    & 0.6945                                &
\\ \cline{1-4}
TCM                                   & 0.7060                                   & 0.6940                                    & 0.6950                                &                            \\ \cline{1-4}
GIN                                 & \textbf{0.7196}                          & \textbf{0.8188}                           & \textbf{0.7692}                       &                            \\ \hline \hline
w-TCM                                 & 0.2980                                   & 0.2410                                    & 0.2695                                & \multirow{3}{*}{Ch-Wiki}   \\ \cline{1-4}
c-TCM                                 & 0.3170                                   & 0.3100                                    & 0.3135                                &                            \\ \cline{1-4}
GIN                                 & \textbf{0.3847}                          & \textbf{0.3343}                           & \textbf{0.3595}                       &                            \\ \hline
\end{tabular}
\end{table}

First, GIN outperforms all the compared methods over the five datasets for the text-query-image task. On the Eng-Wiki
and Pascal datasets, the MAP scores of GIN are 76.72\% and 45.15\%, which are about 35.70\% and 17.14\% higher
than the second best result from JFSSL. For the NUS-WIDE dataset, the MAP score of GIN is 54.18\% and 12.9\% higher than the
second best result from AUSL. It's obvious that no matter for the rich text, e.g. Eng-Wiki and TVGraz, or for the sparse tags,
e.g. NUS-WIDE and Pascal, our model gains the superior performance for the text-query-image task. The reason is that the
proposed model effectively keeps the inter-word semantic relations by representing the texts with graphs, which has been
ignored by other methods that represent the texts with only feature vectors, no mater skip-gram vectors or word frequency vectors.
Such inter-word relations are enhanced and more semantically relevant words are activated with the successive layers
of graph convolutions, resulting in discriminative representations of the text modality.

Second, the MAP score of GIN for the image-query-text task is superior to most of the compared methods. GIN ranks the second
best on Eng-Wiki and NUS-WIDE, the third best on Pascal and the best on TVGraz and Chi-Wiki. Table \ref{tab:MAP} indicates that GIN is only
inferior to JFSSL by 1.44\% on Eng-Wiki and 4.37\% on Pascal. GIN is just 4.54\% lower than AUSL on NUS-WIDE. Since GIN uses
off-the-shelf feature vectors for image view, it's normal that the performance is comparable with state-of-the-art results. The retrieval
performance can be further improved if the feature extraction network was trained together with the fully connected layers in our model.
In this paper, we didn't focus on the vector feature selection problem.

\begin{figure}[!h]
\begin{minipage}{0.45\linewidth}
  \centerline{\includegraphics[width=4.3cm,height = 3.8cm]{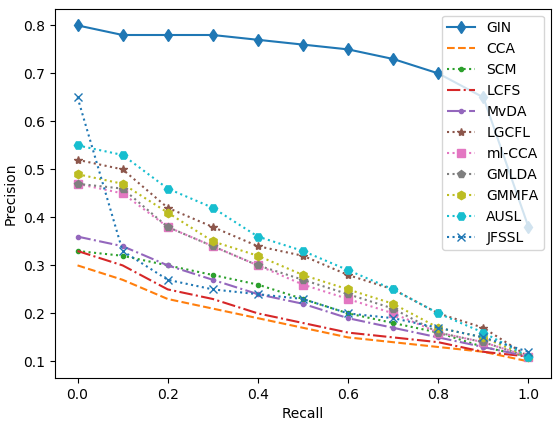}}
  \centerline{\scriptsize{(a) Eng-Wiki: Text-query-images}}
\end{minipage}
\hfill
\begin{minipage}{0.5\linewidth}
  \centerline{\includegraphics[width=4.3cm,height = 3.8cm]{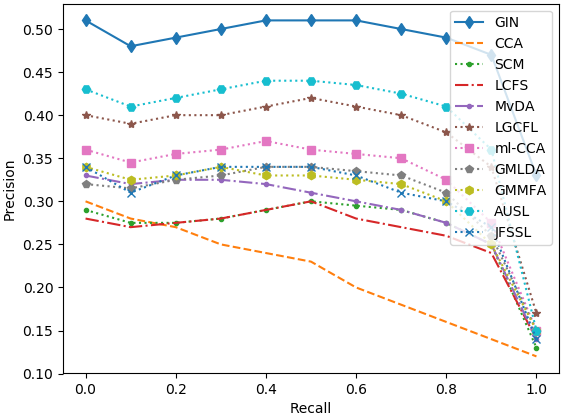}}
  \centerline{\scriptsize{(b) Eng-Wiki: Image-query-texts}}
\end{minipage}

\begin{minipage}{0.45\linewidth}
  \centerline{\includegraphics[width=4.3cm,height = 3.8cm]{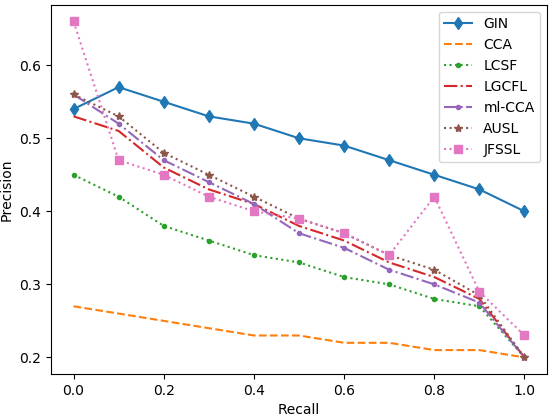}}
  \centerline{\scriptsize{(c) NUS-WIDE: Text-query-images}}
\end{minipage}
\hfill
\begin{minipage}{0.5\linewidth}
  \centerline{\includegraphics[width=4.3cm,height = 3.8cm]{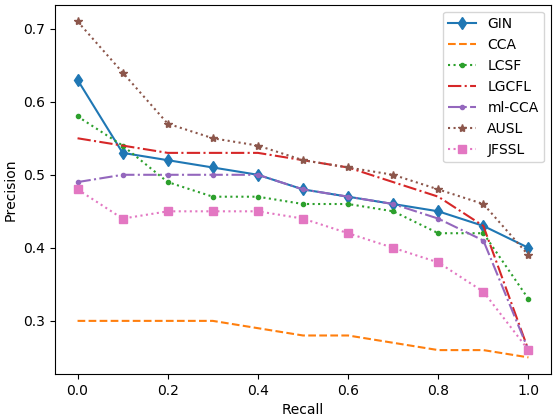}}
  \centerline{\scriptsize{(d) NUS-WIDE: Image-query-texts}}
\end{minipage}

\begin{minipage}{0.45\linewidth}
  \centerline{\includegraphics[width=4.3cm,height = 3.8cm]{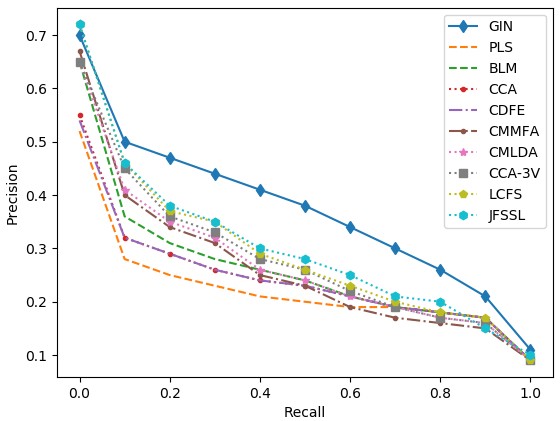}}
  \centerline{\scriptsize{(e) Pascal: Text-query-images}}
\end{minipage}
\hfill
\begin{minipage}{0.5\linewidth}
  \centerline{\includegraphics[width=4.3cm,height = 3.8cm]{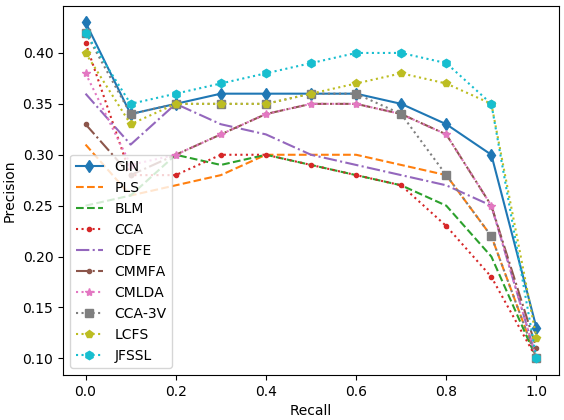}}
  \centerline{\scriptsize{(f) Pascal: Image-query-texts}}
\end{minipage}

\begin{minipage}{0.45\linewidth}
  \centerline{\includegraphics[width=4.3cm,height = 3.8cm]{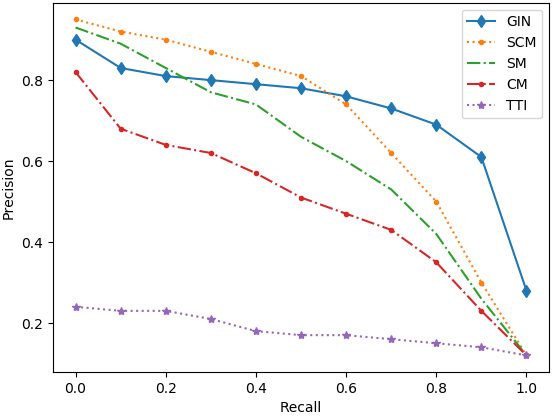}}
  \centerline{\scriptsize{(g) TVGraz: Text-query-images}}
\end{minipage}
\hfill
\begin{minipage}{0.5\linewidth}
  \centerline{\includegraphics[width=4.3cm,height = 3.8cm]{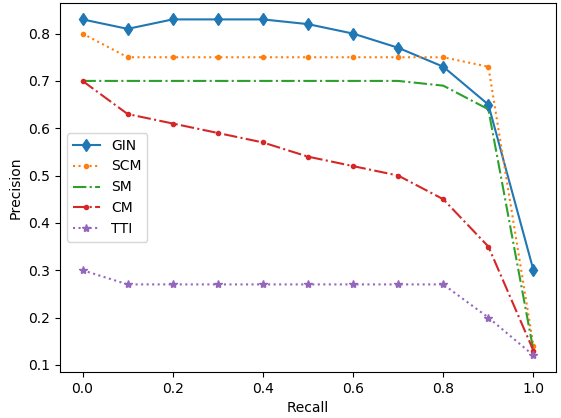}}
  \centerline{\scriptsize{(h) TVGraz: Image-query-texts}}
\end{minipage}

\begin{minipage}{0.45\linewidth}
  \centerline{\includegraphics[width=4.3cm,height = 3.8cm]{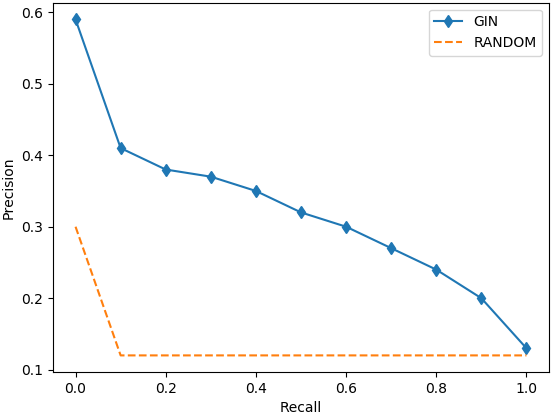}}
  \centerline{\scriptsize{(i) Ch-Wiki: Text-query-images}}
\end{minipage}
\hfill
\begin{minipage}{0.5\linewidth}
  \centerline{\includegraphics[width=4.3cm,height = 3.8cm]{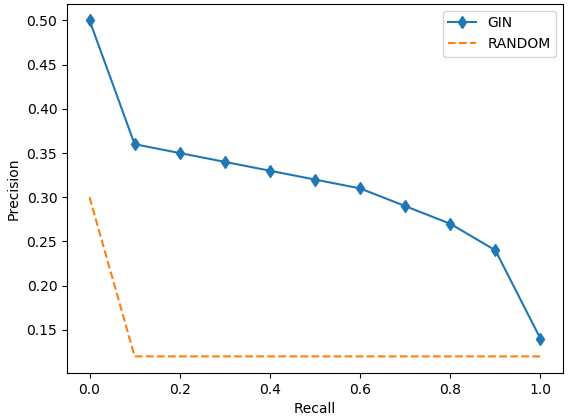}}
  \centerline{\scriptsize{(j) Ch-Wiki: Image-query-texts}}
\end{minipage}

\caption{Precision-recall curves on the five datasets.}
\label{fig: PRcurve}
\end{figure}

Third, GIN achieves the best average MAP over all the competitors, especially outperforming the second best method JFSSL by 17.13\% on Eng-Wiki.
That's mainly because that our learning framework can jointly seek a common latent semantic space and correlated feature representations of multi-modal
data, which can be trained end-to-end. The parameters in the path of graph convolutional networks are learnt referring to the features in the
image branch, which enhances the relations between different modal features in their original data domain. Moreover, the learnt distance
metric is also improving the separation between matching and non-matching image-text pairs.

Finally, on TVGraz dataset, GIN obtains the best results for both retrieval tasks. The improvement for the image-query-text task is
greater than that for the text-query-image task, which is quite different from the observations on other datasets. The reason is that, for
the image view, the existing algorithms represent images simply  by bag-of-features with SIFT descriptors while we utilize the 4096-dimensional
CNN features, which are proved to be much more powerful than the hand-crafted feature descriptors. In addition to English, the representative
alphabetic language, we also conduct experiments on Chinese dataset to show the generalization ability of our model. On
Ch-Wiki, GIN gains 6.77\% and 2.43\% improvement for the text query and image query, respectively.

The precision-recall (PR) curves of image-query-text and text-query-image are plotted
in Figure \ref{fig: PRcurve}. Since the competitive models, i.e. w-TCM and c-TCM, haven't reported PR curves on Ch-Wiki,
 we compare GIN with random baseline on this dataset. For JFSSL,
we show its best MAP after feature selection (see Table 7 in \cite{Wang:joint}).
Since JFSSL hasn't reported the PR curves corresponding to the best MAP, we use its reported PR
curves in \cite{Wang:joint}.

For the text-query-image task, it's obvious that GIN achieves the highest precision than the compared methods with almost all the
recall rate on the five benchmark datasets. For the image-query-text task, GIN outperforms other
competitors with almost all the recall rate on Eng-Wiki. For NUS-WIDE dataset, GIN is only inferior to AUSL
and LGCFL. For Pascal dataset, GIN is just slightly inferior to JFSSL. On the whole, GIN is comparable
with state-of-the-art methods for the image-query-text task.

\textbf{Discussion.} In general, the proposed model shows superior performance for the text-query-image task in all the comparison
experiments, especially on the three widely used benchmark datasets (i.e. Eng-Wiki, NUSE-WIDE, and Pascal), achieving about 17\%$\sim$35\% remarkable improvement
on MAP. It's mainly because that the graph representation for text can well reserve the inherent property of semantic relations between
different words, which provides an effective global prior knowledge for the successive GCN to model each text with distinctive feature input.
In the graph convolutional procedure, such global prior knowledge guides the convolution to adaptively propagate the distinctive vertex features along
semantic paths, which guarantees good generalization ability at conceptual level for the learnt text representation. The remarkable performance
on text query proves that, compared with \textit{vector}-\textit{space} models, the incorporation of semantic structure is a great benefit
and gains better generalization ability for un-seen data.

\section{Conclusion}
\label{sec: conclusion}
In this paper, we propose a novel cross-modal retrieval model named GIN that takes both
irregular graph-structured textual representations and regular vector-structured visual representaions  into consideration to
jointly learn coupled feature and common latent semantic space. A dual path neural network with
graph convolutional networks and layers of nonlinearities is trained using a pairwise
similarity loss function. Extensive experiments on five benchmark datasets
demonstrate that our model considerably outperform the state-of-the-art models. Besides, our model
can be widely used in analyzing heterogeneous data lying on irregular or non-Euclidean domains.

\bibliographystyle{named}
\bibliography{ijcai17}

\end{document}